%% file: gandalf2010.tex
\documentclass[copyright]{eptcs}

\providecommand{\event}{GANDALF 2010}
\providecommand{\volume}{??}
\providecommand{\anno}{2010}
\providecommand{\firstpage}{1}
\providecommand{\eid}{??}

\usepackage{amssymb}
\usepackage{epsf}
\input{Commands}

\input{Environments}

\title{On the Expressiveness of Markovian Process Calculi \\
       with Durational and Durationless Actions}

\author{Marco Bernardo
\institute{Dipartimento di Matematica, Fisica e Informatica --
	   Universit\`a di Urbino ``Carlo Bo'' -- Italy}}

\def\titlerunning{On the Expressiveness of Markovian Process Calculi with Durational and Durationless
		  Actions}
\def\authorrunning{M.\ Bernardo}

\begin{document}

\maketitle

%%%%%%%%%%%%%%%%%%%%%%%%%%%%%%%%%%%%%%%%%%%%%%%%%%%%%%%%%%%%%%%%%
%                                                               %
%                                                               %
% Abstract                                                      %
%                                                               %
%                                                               %
%%%%%%%%%%%%%%%%%%%%%%%%%%%%%%%%%%%%%%%%%%%%%%%%%%%%%%%%%%%%%%%%%
\begin{abstract}

\noindent
Several Markovian process calculi have been proposed in the literature, which differ from each other for
various aspects. With regard to the action representation, we distinguish between integrated-time Markovian
process calculi, in which every action has an exponentially distributed duration associated with it, and
orthogonal-time Markovian process calculi, in which action execution is separated from time passing. Similar
to deterministically timed process calculi, we show that these two options are not irreconcilable by
exhibiting three mappings from an integrated-time Markovian process calculus to an orthogonal-time Markovian
process calculus that preserve the behavioral equivalence of process terms under different interpretations
of action execution: eagerness, laziness, and maximal progress. The mappings are limited to classes of
process terms of the integrated-time Markovian process calculus with restrictions on parallel composition
and do not involve the full capability of the orthogonal-time Markovian process calculus of expressing
nondeterministic choices, thus elucidating the only two important differences between the two calculi: their
synchronization disciplines and their ways of solving choices.

\end{abstract}

%%%%%%%%%%%%%%%%%%%%%%%%%%%%%%%%%%%%%%%%%%%%%%%%%%%%%%%%%%%%%%%%%
%
%
\section{Introduction}\label{intro}
%
%
%%%%%%%%%%%%%%%%%%%%%%%%%%%%%%%%%%%%%%%%%%%%%%%%%%%%%%%%%%%%%%%%%

Communicating concurrent systems are characterized not only by their functional behavior, but also by their
quantitative features. A prominent role is played by timing aspects, which express the temporal ordering of
system activities and are of paramount importance in the study of the properties of real-time systems as
well as shared-resource systems. As witnessed by a rich literature, there are several different options for
introducing time and time passing in system descriptions, many of which have been formalized in a process
algebraic setting~\cite{ABC}.

Starting from the late 80's, a number of deterministically timed process calculi have been proposed -- like,
e.g., timed CSP~\cite{RR}, temporal CCS~\cite{MT}, timed CCS~\cite{Yi}, real-time ACP~\cite{BB}, urgent
LOTOS~\cite{BL}, TIC~\cite{QFA}, ATP~\cite{NS2}, TPL~\cite{HRe}, \textsc{cIpa}~\cite{AM}, and
PAFAS~\cite{CVJ} -- in which time and time passing are represented through a dense or discrete time domain
-- like, e.g., $(\natns, +, \le)$ -- equipped with an associative operation with neutral element and a total
order defined on the basis of this operation. As observed in~\cite{NS1,UY,Cor}, the various
deterministically timed process calculi differ for a number of time-related options, some of which give rise
to the one-phase functioning principle -- according to which actions are durational, time is absolute, and
several local clocks are present -- and the two-phase functioning principle -- according to which actions
are durationless, time is relative, and there is a single global clock:

	\begin{itemize}

\item Durational actions versus durationless actions. In the first case, every action takes a fixed amount
of time to be performed and time passes only due to action execution; hence, functional behavior and time
passing are integrated. In the second case, actions are instantaneous events and time passes in between
them; hence, functional behavior and time passing are orthogonal.

\item Absolute time versus relative time. Assuming that timestamps are associated with the events observed
during system execution, in the first case all timestamps refer to the starting time of the system
execution, while in the second case each timestamp refers to the starting time or the completion time of the
previously observed event (the two times coincide if events are durationless).

\item Local clocks versus global clock. In the first case, there are several clocks associated with the
various system parts, which elapse independent of each other although they define a unique notion of global
time. In the second case, there is a single clock that governs time passing.

	\end{itemize}

Another degree of freedom is concerned with the different interpretations of action execution, in terms of
whether and when it can be delayed. There are at least the following three interpretations:

	\begin{itemize}

\item Eagerness, which establishes that actions must be performed as soon as they become enabled without any
delay, thereby implying that actions are urgent.

\item Laziness, which establishes that, once they become enabled, actions can be delayed arbitrarily long
before they are executed.

\item Maximal progress, which establishes that actions can be delayed arbitrarily long unless they are
involved in synchronizations, in which case they are urgent.

	\end{itemize}

In~\cite{Cor}, a translating function is defined from a one-phase deterministically timed process calculus
inspired by~\cite{AM} to a two-phase deterministically timed process calculus inspired by~\cite{MT}, which
is shown to preserve the behavioral equivalence of process terms based on CCS-like parallel composition and
restriction operators~\cite{Mil}. The result holds under eagerness (only for restriction-free terms),
laziness, and maximal progress, both when observing the starting time of action execution and when observing
the completion time of action execution. This demonstrates that the different choices that can be made about
the representation of time and time passing in a deterministically timed framework are not irreconcilable.

Starting from the first half of the 90's, a number of stochastically timed process calculi have been
proposed too -- like, e.g., TIPP~\cite{GHR,HR}, PEPA~\cite{Hil}, MPA~\cite{Buc}, \empagr~\cite{BBr},
S$\pi$~\cite{Pri}, IMC~\cite{Her}, and PIOA~\cite{SCS} -- in which time and time passing are represented by
means of exponentially distributed random variables rather than nonnegative numbers. The reason for using
only exponential distributions (uniquely identified through their rates, positive real numbers corresponding
to the reciprocal of their expected values) is twofold. Firstly, the stochastic process underlying a system
description turns out to be a continuous-time Markov chain, which simplifies quantitative analysis without
sacrificing expressiveness. Secondly, the memoryless property of exponential distributions fits well with
the interleaving view of concurrency.

The time-related options and the action execution interpretations discussed for deterministically timed
process calculi apply to a large extent also to stochastically timed process calculi. This is especially
true for the difference between durational actions and durationless actions, which results in
integrated-time Markovian process calculi like TIPP, PEPA, MPA, \empagr, S$\pi$, and PIOA and
orthogonal-time Markovian process calculi like IMC, respectively. By contrast, the distinction between
absolute time and relative time and the concept of clock are not important in a Markovian framework. Due to
the memoryless property of exponential distributions, only rates of durational actions or time delays
matter.

A remarkable difference between deterministically timed process calculi and stochastically timed process
calculi is concerned with the way choices among alternative behaviors are solved. In the first case, the
choice is nondeterministic precisely as in classical process calculi, which means that time does not solve
choices. In an orthogonal-time setting, this is witnessed by the presence of operational semantic rules
according to which a process term of the form $(n) \, . \, Q_{1} + (n) \, . \, Q_{2}$ -- where $+$ denotes
the alternative composition operator -- can let $n$ time units pass and then evolves into $Q_{1} + Q_{2}$.
In the second case, the choice can instead be probabilistic whenever exponentially distributed delays come
into play. In the same orthogonal-time setting, a process term of the form $(\lambda_{1}) \, . \, Q_{1} +
(\lambda_{2}) \, . \, Q_{2}$ -- where $\lambda_{1}$ and $\lambda_{2}$ are the rates of two exponentially
distributed delays -- evolves into $Q_{1}$ or $Q_{2}$ with probabilities ${\lambda_{1} \over \lambda_{1} +
\lambda_{2}}$ and ${\lambda_{2} \over \lambda_{1} + \lambda_{2}}$.

This has an impact on the expressiveness of Markovian process calculi. In fact, the orthogonal-time ones are
more expressive than the integrated-time ones, because the former can represent both action-based
nondeterministic choices and time-based probabilistic choices, whereas the latter can represent only
probabilistic choices based on action durations. In turn, this has an impact on the expressiveness of the
synchronization discipline adopted in the considered calculi. In fact, in the orthogonal-time case the time
to the synchronization of two actions can be naturally expressed as the maximum of two exponentially
distributed delays, whereas in the integrated-time case the duration of the synchronization of two
exponentially timed actions has to be assumed to be exponentially distributed with rate given by the
application of an associative and commutative operation to the two original rates.

Another important difference between deterministically timed process calculi and stochastically timed
process calculi is concerned with the formalization of the various interpretations of action execution. In
the first case, all the three interpretations can be encoded in the operational semantic rules. In the
second case, it depends on whether time is integrated with action execution or separated from it, as we show
in this paper. On the one hand, observed that the usual operational semantic rules for integrated-time
Markovian process calculi encode eagerness as they permit no delay, we recognize that the same rules encode
laziness and maximal progress too, because the possibility of delaying the beginning of action execution is
inherent in the memoryless property of exponentially distributed durations. On the other hand, since
additional operational semantic rules delaying action execution would produce no effect in orthogonal-time
Markovian process calculi as time can solve choices, we exploit the behavioral equivalence to express when
action execution takes precedence over time passing.

In spite of the different expressiveness they induce, in this paper we show that durational actions and
durationless actions are not irreconcilable even in a Markovian setting. Similar to~\cite{Cor}, this is
accomplished by defining three translating functions from an integrated-time Markovian process calculus to
an orthogonal-time Markovian process calculus that preserve the behavioral equivalence of process terms
under eagerness, laziness, and maximal progress, respectively. The encodings are limited to classes of
process terms of the integrated-time Markovian process calculus with restrictions on parallel composition
and do not involve the full capability of the orthogonal-time Markovian process calculus of expressing
nondeterministic choices. This formally clarifies the only two important differences between the two
calculi, i.e., their different synchronization disciplines and their different ways of solving choices.

This paper is organized as follows. In Sects.~\ref{mpcdur} and~\ref{mpcnodur}, we uniformly present the
syntax, the operational semantics, and a bisimulation-based behavioral equivalence for an integrated-time
Markovian process calculus and an orthogonal-time Markovian process calculus, respectively, and we discuss
how to represent the three different interpretations of action execution. Then, in Sect.~\ref{encoding} we
exhibit the three encodings from certain classes of process terms of the integrated-time Markovian process
calculus to certain classes of process terms of the orthogonal-time Markovian process calculus and we
demonstrate that they preserve the bisimulation-based behavioral equivalence of the considered process
terms. Finally, in Sect.~\ref{concl} we report some concluding remarks.

%%%%%%%%%%%%%%%%%%%%%%%%%%%%%%%%%%%%%%%%%%%%%%%%%%%%%%%%%%%%%%%%%
%
%
\section{Markovian Process Calculus with Durational Actions}\label{mpcdur}
%
%
%%%%%%%%%%%%%%%%%%%%%%%%%%%%%%%%%%%%%%%%%%%%%%%%%%%%%%%%%%%%%%%%%

In this section, we present a Markovian process calculus inspired by~\cite{Hil,HR} in which every action has
associated with it a rate that uniquely identifies its exponentially distributed duration. The presentation
of the integrated-time Markovian process calculus -- ITMPC for short -- consists of the definition of its
syntax, its operational semantics, and a bisimulation-based behavioral equivalence. A discussion of the
interpretation of action execution accompanies the definition of the operational semantics.

%%%%%%%%%%%%%%%%%%%%%%%%%%%%%%%%%%%%%%%%%%%%%%%%%%%%%%%%%%%%%%%%%
%
\subsection{Durational Actions and Behavioral Operators}
%
%%%%%%%%%%%%%%%%%%%%%%%%%%%%%%%%%%%%%%%%%%%%%%%%%%%%%%%%%%%%%%%%%

In ITMPC, an exponentially timed action is represented as a pair $\lap a, \lambda \rap$. The first element,
$a$, is the name of the action, which is $\tau$ in the case that the action is internal, otherwise it
belongs to a set $\ms{Name}_{\rm v}$ of visible action names. The second element, $\lambda \in \realns_{>
0}$, is the rate of the exponentially distributed random variable $\ms{RV}$ quantifying the duration of the
action, i.e., $\Pr \{ \ms{RV} \le t \} = 1 - {\rm e}^{- \lambda \cdot t}$ for $t \in \realns_{> 0}$. The
average duration of the action is equal to the reciprocal of its rate, i.e., $1 / \lambda$. If several
exponentially timed actions are enabled, the race policy is adopted: the action that is executed is the
fastest one.

The sojourn time associated with a process term $P$ is thus the minimum of the random variables quantifying
the durations of the exponentially timed actions enabled by $P$. Since the minimum of several exponentially
distributed random variables is exponentially distributed and its rate is the sum of the rates of the
original variables, the sojourn time associated with $P$ is exponentially distributed with rate equal to the
sum of the rates of the actions enabled by $P$. Therefore, the average sojourn time associated with~$P$ is
the reciprocal of the sum of the rates of the actions it enables. The probability of executing one of those
actions is given by the action rate divided by the sum of the rates of all the considered actions.

ITMPC comprises a CSP-like parallel composition operator~\cite{Hoa} according to which two exponentially
timed actions synchronize iff they have the same visible name belonging to an explicit synchronization set.
The resulting action has the same name as the two original actions and its rate is obtained by applying an
associative and commutative operator $\otimes$ to the rates of the two original actions.

We denote by $\ms{Act}_{\rm M, it} = \ms{Name} \times \realns_{> 0}$ the set of actions of ITMPC, where
$\ms{Name} = \ms{Name}_{\rm v} \cup \{ \tau \}$ is the set of action names -- ranged over by $a, b$ -- and
$\realns_{> 0}$ is the set of action rates -- ranged over by $\lambda, \mu$. We then denote by $\ms{Relab}$
a set of relabeling functions $\varphi : \ms{Name} \rightarrow \ms{Name}$ that preserve action visibility,
i.e., such that $\varphi^{-1}(\tau) = \{ \tau \}$. Finally, we denote by $\ms{Var}$ a set of process
variables ranged over by $X, Y, Z$.

	\begin{definition}

The process language $\calpl_{\rm M, it}$ is generated by the following syntax:
\[\begin{array}{|rcll|}
\hline
P & \!\!\! ::= \!\!\! & \nil & \hspace{0.5cm} \textrm{inactive process} \\
& \!\!\! | \!\!\! & \lap a, \lambda \rap . P & \hspace{0.5cm} \textrm{exponentially timed action prefix} \\
& \!\!\! | \!\!\! & P + P & \hspace{0.5cm} \textrm{alternative composition} \\
& \!\!\! | \!\!\! & P \pco{S} P & \hspace{0.5cm} \textrm{parallel composition} \\
& \!\!\! | \!\!\! & P / H & \hspace{0.5cm} \textrm{hiding} \\
& \!\!\! | \!\!\! & P[\varphi] & \hspace{0.5cm} \textrm{relabeling} \\
& \!\!\! | \!\!\! & X & \hspace{0.5cm} \textrm{process variable} \\
& \!\!\! | \!\!\! & \textrm{rec} \, X : P & \hspace{0.5cm} \textrm{recursion} \\
\hline
\end{array}\]
where $a \in \ms{Name}$, $\lambda \in \realns_{> 0}$, $S, H \subseteq \ms{Name}_{\rm v}$, $\varphi \in
\ms{Relab}$, and $X \in \ms{Var}$. We denote by $\procs_{\rm M, it}$ the set of closed and guarded process
terms of $\calpl_{\rm M, it}$.
\fullbox

	\end{definition}

%%%%%%%%%%%%%%%%%%%%%%%%%%%%%%%%%%%%%%%%%%%%%%%%%%%%%%%%%%%%%%%%%
%
\subsection{Integrated-Time Operational Semantics: Eagerness, Laziness, Maximal Progress}\label{discussit}
%
%%%%%%%%%%%%%%%%%%%%%%%%%%%%%%%%%%%%%%%%%%%%%%%%%%%%%%%%%%%%%%%%%

The semantics for ITMPC can be defined in the usual operational style, with an important difference with
respect to the nondeterministic case. A process term like $\lap a, \lambda \rap . \nil + \lap a, \lambda
\rap . \nil$ is not the same as $\lap a, \lambda \rap . \nil$, because the average sojourn time associated
with the latter, i.e., $1 / \lambda$, is twice the average sojourn time associated with the former, i.e., $1
/ (\lambda + \lambda)$. A way of assigning distinct semantic models to terms like the two considered above
consists of taking into account the multiplicity of each transition, intended as the number of different
proofs for the transition derivation.

\newpage

The semantic model $\lsp P \rsp_{\rm M, it}$ for a process term $P \in \procs_{\rm M, it}$ is thus a labeled
multitransition system. Its multitransition relation is contained in the smallest multiset of elements of
$\procs_{\rm M, it} \times \ms{Act}_{\rm M, it} \times \procs_{\rm M, it}$ that satisfy the operational
semantic rules of Table~\ref{sosit} -- where $\{ \_ \hookrightarrow \_ \}$ denotes syntactical replacement
-- and keep track of all the possible ways of deriving each transition.

	\begin{table}[thb]

\[\begin{array}{|c|}
\hline
(\textsc{Pre}_{\rm M, it}) \quad {\infr{}{\lap a, \lambda \rap . P \arrow{a, \lambda}{\rm M, it} P}}
\\[0.9cm]
(\textsc{Alt}_{\rm M, it, 1}) \quad {\infr{P_{1} \arrow{a, \lambda}{\rm M, it} P'}{P_{1} + P_{2} \arrow{a,
\lambda}{\rm M, it} P'}} \hspace{1.5cm}
(\textsc{Alt}_{\rm M, it, 2}) \quad {\infr{P_{2} \arrow{a, \lambda}{\rm M, it} P'}{P_{1} + P_{2} \arrow{a,
\lambda}{\rm M, it} P'}} \\[0.9cm]
(\textsc{Par}_{\rm M, it, 1}) \quad {\infr{P_{1} \arrow{a, \lambda}{\rm M, it} P'_{1} \hspace{0.8cm} a
\notin S}{P_{1} \pco{S} P_{2} \arrow{a, \lambda}{\rm M, it} P'_{1} \pco{S} P_{2}}} \hspace{1.5cm}
(\textsc{Par}_{\rm M, it, 2}) \quad {\infr{P_{2} \arrow{a, \lambda}{\rm M, it} P'_{2} \hspace{0.8cm} a
\notin S}{P_{1} \pco{S} P_{2} \arrow{a, \lambda}{\rm M, it} P_{1} \pco{S} P'_{2}}} \\[0.9cm]
(\textsc{Syn}_{\rm M, it}) \quad {\infr{P_{1} \arrow{a, \lambda_{1}}{\rm M, it} P'_{1} \hspace{0.8cm} P_{2}
\arrow{a, \lambda_{2}}{\rm M, it} P'_{2} \hspace{0.8cm} a \in S}{P_{1} \pco{S} P_{2} \arrow{a, \lambda_{1}
\otimes \lambda_{2}}{\rm M, it} P'_{1} \pco{S} P'_{2}}} \\[0.9cm]
(\textsc{Hid}_{\rm M, it, 1}) \quad {\infr{P \arrow{a, \lambda}{\rm M, it} P' \hspace{0.8cm} a \in H}{P / H
\arrow{\tau, \lambda}{\rm M, it} P' / H}} \hspace{1.5cm}
(\textsc{Hid}_{\rm M, it, 2}) \quad {\infr{P \arrow{a, \lambda}{\rm M, it} P' \hspace{0.8cm} a \notin H}{P /
H \arrow{a, \lambda}{\rm M, it} P' / H}} \\[0.9cm]
\hspace*{1.6cm} (\textsc{Rel}_{\rm M, it}) \quad {\infr{P \arrow{a, \lambda}{\rm M, it} P'}{P[\varphi]
\arrow{\varphi(a), \lambda}{\rm M, it} P'[\varphi]}} \hspace{1.5cm}
(\textsc{Rec}_{\rm M, it}) \quad {\infr{P \{ \textrm{rec} \, X : P \hookrightarrow X \} \arrow{a,
\lambda}{\rm M, it} P'}{\textrm{rec} \, X : P \arrow{a, \lambda}{\rm M, it} P'}} \\
\hline
\end{array}\]

\caption{Operational semantic rules for ITMPC}\label{sosit}

	\end{table}

These operational semantic rules encode an eager interpretation of action execution, as they permit no delay
between the time at which an exponentially timed action becomes enabled and the time at which the same
action starts its execution. This is the standard interpretation adopted by all the integrated-time
Markovian process calculi appeared in the literature. However, the operational semantic rules of
Table~\ref{sosit} encode laziness and maximal progress too, because the possibility of delaying the
beginning of action execution is inherent in the memoryless property of exponentially distributed durations.
In fact, if an exponentially timed action does not finish its execution within $t$ time units, the residual
execution time has the same distribution as the whole action duration and thus the beginning of the
execution of the action can be thought of as being delayed by $t$ time units with respect to the instant in
which the action has become enabled.

\newpage

Recalling that in the durational setting defined in~\cite{Cor} every state is a pair $k \Rightarrow P$ where
$k$ is the clock and $P$ is the process, the operational semantic rule for lazy deterministically timed
actions is of the form:
\cws{0}{k \Rightarrow \lap a, n \rap . P \arrow{a, n}{} (k + t + n) \Rightarrow P \hspace{0.5cm} \forall t
\in \natns}
where $k \in \natns$ is the value of the clock when the action becomes enabled, $n \in \natns$ is the fixed
duration of the action, and $k + t + n$ is the value of the clock when the action finishes its execution,
with $t$ being an arbitrary delay between the time at which the action becomes enabled and the time at which
the action starts its execution. In the maximal progress case, the above rule is applied only when $a \in
\ms{Name}_{\rm v}$, while the rule for deterministically timed $\tau$-actions is still of the form:
\cws{0}{k \Rightarrow \lap \tau, n \rap . P \arrow{\tau, n}{} (k + n) \Rightarrow P}
which enforces an eager interpretation of those actions because they cannot be delayed.

Since in a Markovian framework it is possible to express rates but not fixed durations, the only analogous
operational semantic rule for lazy exponentially timed actions would be of the form:
\cws{0}{\lap a, \lambda \rap . P \arrow{a, \lambda'}{} P \hspace{0.5cm} \forall \lambda' \in \realns_{]0,
\lambda]}}
However, this would represent an action slowdown rather than delaying the beginning of the action execution
by an arbitrary amount of time and then performing the action at its rate. An appropriate semantic treatment
of lazy exponentially timed actions should not alter their rates. Therefore, a better option is to add a
further operational semantic rule for action prefix of the form:
\cws{0}{\lap a, \lambda \rap . P \arrow{\tau, \lambda'}{} \lap a, \lambda \rap . P \hspace{0.5cm} \forall
\lambda' \in \realns_{> 0}}
which introduces invisible selfloops each having an arbitrary rate. But these selfloops have no impact on
(the transient/stationary state probabilities of) the underlying continuous-time Markov chain. In fact,
thanks to the memoryless property of exponential distributions, the time remaining to moving from $\lap a,
\lambda \rap . P$ to~$P$ after the execution of an arbitrary number of selfloops is still exponentially
distributed with rate~$\lambda$. As a consequence, the introduction of these selfloops is useless, which
means that the operational semantic rules of Table~\ref{sosit} encode also laziness. Since maximal progress
is in some sense between eagerness and laziness, it is encoded in those rules as well.

%%%%%%%%%%%%%%%%%%%%%%%%%%%%%%%%%%%%%%%%%%%%%%%%%%%%%%%%%%%%%%%%%
%
\subsection{Integrated-Time Markovian Bisimilarity}
%
%%%%%%%%%%%%%%%%%%%%%%%%%%%%%%%%%%%%%%%%%%%%%%%%%%%%%%%%%%%%%%%%%

A behavioral equivalence over $\procs_{\rm M, it}$ can be defined by establishing that, whenever a process
term can perform actions with a certain name that reach a certain set of terms at a certain speed, then any
process term equivalent to the given one has to be able to respond with actions with the same name that
reach an equivalent set of terms at the same speed. This can be easily formalized through the comparison of
the process term exit rates.

The integrated-time exit rate of a process term $P \in \procs_{\rm M, it}$ is the rate at which $P$ can
execute actions of a certain name $a \in \ms{Name}$ that lead to a certain destination $D \subseteq
\procs_{\rm M, it}$ and is given by the sum of the rates of those actions due to the race policy:
\[\begin{array}{|c|}
\hline
\ms{rate}_{\rm it}(P, a, D) \: = \: \sum \lmp \lambda \in \realns_{> 0} \mid \exists P' \in D \ldotp P
\arrow{a, \lambda}{\rm M, it} P' \rmp \\
\hline
\end{array}\]
where $\lmp$ and $\rmp$ are multiset delimiters and the summation is taken to be zero if its multiset is
empty. By summing up the rates of all the actions of $P$, we obtain the integrated-time total exit rate of
$P$:
\[\begin{array}{|c|}
\hline
\ms{rate}_{\rm it, t}(P) \: = \: \sum\limits_{a \in \ms{Name}} \ms{rate}_{\rm it}(P, a, \procs_{\rm M, it})
\\
\hline
\end{array}\]
which coincides with the reciprocal of the average sojourn time associated with $P$.

	\begin{definition}

An equivalence relation $\calb$ over $\procs_{\rm M, it}$ is an integrated-time Markovian bisimulation iff,
whenever $(P_{1}, P_{2}) \in \calb$, then for all action names $a \in \ms{Name}$ and equivalence classes $D
\in \procs_{\rm M, it} / \calb$:
\cws{0}{\ms{rate}_{\rm it}(P_{1}, a, D) \: = \: \ms{rate}_{\rm it}(P_{2}, a, D)}
Integrated-time Markovian bisimilarity $\sbis{\rm MB, it}$ is the union of all the integrated-time Markovian
bisimulations.
\fullbox

	\end{definition}

$\sbis{\rm MB, it}$ can be shown to be a congruence with respect to all the operators of ITMPC as well as
recursion, and to have a sound and complete axiomatization over nonrecursive process terms including typical
laws like associativity, commutativity, and neutral element for the alternative composition operator, the
expansion law for the parallel composition operator, and distributive laws for hiding and relabeling with
respect to alternative composition. Its characterizing law -- which replaces the usual idempotency of the
alternative composition operator and encodes the race policy -- is the following:
\cws{12}{\lap a, \lambda_{1} \rap . P + \lap a, \lambda_{2} \rap . P \: \sbis{\rm MB, it} \: \lap a,
\lambda_{1} + \lambda_{2} \rap . P}

%%%%%%%%%%%%%%%%%%%%%%%%%%%%%%%%%%%%%%%%%%%%%%%%%%%%%%%%%%%%%%%%%
%
%
\section{Markovian Process Calculus with Durationless Actions}\label{mpcnodur}
%
%
%%%%%%%%%%%%%%%%%%%%%%%%%%%%%%%%%%%%%%%%%%%%%%%%%%%%%%%%%%%%%%%%%

In this section, we present a Markovian process calculus inspired by~\cite{Her} in which actions are
durationless and hence action execution is separated from time passing. The presentation of the
orthogonal-time Markovian process calculus -- OTMPC for short -- consists of the definition of its syntax,
its operational semantics, and a bisimulation-based behavioral equivalence. A discussion of the
interpretation of action execution accompanies the definition of the behavioral equivalence.

%%%%%%%%%%%%%%%%%%%%%%%%%%%%%%%%%%%%%%%%%%%%%%%%%%%%%%%%%%%%%%%%%
%
\subsection{Durationless Actions, Time Passing, and Behavioral Operators}
%
%%%%%%%%%%%%%%%%%%%%%%%%%%%%%%%%%%%%%%%%%%%%%%%%%%%%%%%%%%%%%%%%%

In OTMPC, actions are instantaneous and time passes in between them. As a consequence, there are two prefix
operators: an action prefix operator $a \, . \_$, with $a \in \ms{Name}$, and a time prefix operator
$(\lambda) \, . \_$, with \linebreak $\lambda \in \realns_{> 0}$. Similar to ITMPC, time delays are governed
by exponential distributions and are subject to the race policy. Different from ITMPC, the CSP-like parallel
composition operator enforces synchronizations only between two actions that have the same visible name
belonging to the synchronization set; hence, time delays are not involved in synchronizations. Moreover, the
choice among alternative actions is nondeterministic.

	\begin{definition}

The process language $\calpl_{\rm M, ot}$ is generated by the following syntax:
\[\begin{array}{|rcll|}
\hline
Q & \!\!\! ::= \!\!\! & \nil & \hspace{0.5cm} \textrm{inactive process} \\
& \!\!\! | \!\!\! & a \, . \, Q & \hspace{0.5cm} \textrm{action prefix} \\
& \!\!\! | \!\!\! & (\lambda) \, . \, Q & \hspace{0.5cm} \textrm{time prefix} \\
& \!\!\! | \!\!\! & Q + Q & \hspace{0.5cm} \textrm{alternative composition} \\
& \!\!\! | \!\!\! & Q \pco{S} Q & \hspace{0.5cm} \textrm{parallel composition} \\
& \!\!\! | \!\!\! & Q / H & \hspace{0.5cm} \textrm{hiding} \\
& \!\!\! | \!\!\! & Q[\varphi] & \hspace{0.5cm} \textrm{relabeling} \\
& \!\!\! | \!\!\! & X & \hspace{0.5cm} \textrm{process variable} \\
& \!\!\! | \!\!\! & \textrm{rec} \, X : Q & \hspace{0.5cm} \textrm{recursion} \\
\hline
\end{array}\]
where $a \in \ms{Name}$, $\lambda \in \realns_{> 0}$, $S, H \subseteq \ms{Name}_{\rm v}$, $\varphi \in
\ms{Relab}$, and $X \in \ms{Var}$. We denote by $\procs_{\rm M, ot}$ the set of closed and guarded process
terms of $\calpl_{\rm M, ot}$.
\fullbox

	\end{definition}

	\begin{table}[p]

\[\begin{array}{|c|}
\hline
(\textsc{Pre}) \quad {\infr{}{a \, . \, Q \arrow{a}{} Q}} \\[0.9cm]
(\textsc{Alt}_{1}) \quad {\infr{Q_{1} \arrow{a}{} Q'}{Q_{1} + Q_{2} \arrow{a}{} Q'}} \hspace{1.5cm}
(\textsc{Alt}_{2}) \quad {\infr{Q_{2} \arrow{a}{} Q'}{Q_{1} + Q_{2} \arrow{a}{} Q'}} \\[0.9cm]
(\textsc{Par}_{1}) \quad {\infr{Q_{1} \arrow{a}{} Q'_{1} \hspace{0.8cm} a \notin S}{Q_{1} \pco{S} Q_{2}
\arrow{a}{} Q'_{1} \pco{S} Q_{2}}} \hspace{1.5cm}
(\textsc{Par}_{2}) \quad {\infr{Q_{2} \arrow{a}{} Q'_{2} \hspace{0.8cm} a \notin S}{Q_{1} \pco{S} Q_{2}
\arrow{a}{} Q_{1} \pco{S} Q'_{2}}} \\[0.9cm]
(\textsc{Syn}) \quad {\infr{Q_{1} \arrow{a}{} Q'_{1} \hspace{0.8cm} Q_{2} \arrow{a}{} Q'_{2} \hspace{0.8cm}
a \in S}{Q_{1} \pco{S} Q_{2} \arrow{a}{} Q'_{1} \pco{S} Q'_{2}}} \\[0.9cm]
(\textsc{Hid}_{1}) \quad {\infr{Q \arrow{a}{} Q' \hspace{0.8cm} a \in H}{Q / H \arrow{\tau}{} Q' / H}}
\hspace{1.5cm}
(\textsc{Hid}_{2}) \quad {\infr{Q \arrow{a}{} Q' \hspace{0.8cm} a \notin H}{Q / H \arrow{a}{} Q' / H}}
\\[0.9cm]
\hspace*{1.6cm} (\textsc{Rel}) \quad {\infr{Q \arrow{a}{} Q'}{Q[\varphi] \arrow{\varphi(a)}{} Q'[\varphi]}}
\hspace{1.5cm}
(\textsc{Rec}) \quad {\infr{Q \{ \textrm{rec} \, X : Q \hookrightarrow X \} \arrow{a}{} Q'}{\textrm{rec} \,
X : Q \arrow{a}{} Q'}} \\[1.0cm]
\hline
(\textsc{Pre}_{\rm M}) \quad {\infr{}{(\lambda) \, . \, Q \arrow{\lambda}{\rm M} Q}} \\[0.9cm]
(\textsc{Alt}_{\rm M, 1}) \quad {\infr{Q_{1} \arrow{\lambda}{\rm M} Q'}{Q_{1} + Q_{2} \arrow{\lambda}{\rm M}
Q'}} \hspace{1.5cm}
(\textsc{Alt}_{\rm M, 2}) \quad {\infr{Q_{2} \arrow{\lambda}{\rm M} Q'}{Q_{1} + Q_{2} \arrow{\lambda}{\rm M}
Q'}} \\[0.9cm]
(\textsc{Par}_{\rm M, 1}) \quad {\infr{Q_{1} \arrow{\lambda}{\rm M} Q'_{1}}{Q_{1} \pco{S} Q_{2}
\arrow{\lambda}{\rm M} Q'_{1} \pco{S} Q_{2}}} \hspace{1.5cm}
(\textsc{Par}_{\rm M, 2}) \quad {\infr{Q_{2} \arrow{\lambda}{\rm M} Q'_{2}}{Q_{1} \pco{S} Q_{2}
\arrow{\lambda}{\rm M} Q_{1} \pco{S} Q'_{2}}} \\[0.9cm]
(\textsc{Hid}_{\rm M}) \quad {\infr{Q \arrow{\lambda}{\rm M} Q'}{Q / H \arrow{\lambda}{\rm M} Q' / H}}
\\[0.9cm]
\hspace*{1.6cm} (\textsc{Rel}_{\rm M}) \quad {\infr{Q \arrow{\lambda}{\rm M} Q'}{Q[\varphi]
\arrow{\lambda}{\rm M} Q'[\varphi]}} \hspace{1.5cm}
(\textsc{Rec}_{\rm M}) \quad {\infr{Q \{ \textrm{rec} \, X : Q \hookrightarrow X \} \arrow{\lambda}{\rm M}
Q'}{\textrm{rec} \, X : Q \arrow{\lambda}{\rm M} Q'}} \\[1.0cm]
\hline
\end{array}\]

\caption{Operational semantic rules for OTMPC: action transitions and time transitions}\label{sosot}

	\end{table}

%%%%%%%%%%%%%%%%%%%%%%%%%%%%%%%%%%%%%%%%%%%%%%%%%%%%%%%%%%%%%%%%%
%
\subsection{Orthogonal-Time Operational Semantics}
%
%%%%%%%%%%%%%%%%%%%%%%%%%%%%%%%%%%%%%%%%%%%%%%%%%%%%%%%%%%%%%%%%%

The semantics for OTMPC relies on two transition relations: one for action execution and one for time
passing. Like for nondeterministic processes, the former is defined as the smallest subset of $\procs_{\rm
M, ot} \times \ms{Name} \times \procs_{\rm M, ot}$ satisfying the operational semantic rules in the upper
part of Table~\ref{sosot}. Since $(\lambda) \, . \, Q + (\lambda) \, . \, Q$ is not the same as $(\lambda)
\, . \, Q$, the latter is defined as the smallest multiset of elements of $\procs_{\rm M, ot} \times
\realns_{> 0} \times \procs_{\rm M, ot}$ that satisfy the operational semantic rules in the lower part of
Table~\ref{sosot} and keep track of all the possible ways of deriving each transition. The semantic model
$\lsp Q \rsp_{\rm M, ot}$ for a process term $Q \in \procs_{\rm M, ot}$ is thus a labeled multitransition
system, which can contain both nondeterministic and probabilistic branchings.

%%%%%%%%%%%%%%%%%%%%%%%%%%%%%%%%%%%%%%%%%%%%%%%%%%%%%%%%%%%%%%%%%
%
\subsection{Orthogonal-Time Markovian Bisimilarity: Eagerness, Laziness, Maximal Progress}\label{discussot}
%
%%%%%%%%%%%%%%%%%%%%%%%%%%%%%%%%%%%%%%%%%%%%%%%%%%%%%%%%%%%%%%%%%

A behavioral equivalence over $\procs_{\rm M, ot}$ can be defined by combining classical bisimilarity for
action execution with exit rate comparison for time passing. The orthogonal-time exit rate of a process term
$Q \in \procs_{\rm M, ot}$ is the rate at which $Q$ can let time pass when going to a certain destination $D
\subseteq \procs_{\rm M, ot}$ and is given by the sum of the rates of $Q$ delays leading to $D$ due to the
race policy:
\[\begin{array}{|c|}
\hline
\ms{rate}_{\rm ot}(Q, D) \: = \: \sum \lmp \lambda \in \realns_{> 0} \mid \exists Q' \in D \ldotp Q
\arrow{\lambda}{\rm M} Q' \rmp \\
\hline
\end{array}\]
By summing up the rates of all the delays of $Q$, we obtain the orthogonal-time total exit rate of $Q$:
\[\begin{array}{|c|}
\hline
\ms{rate}_{\rm ot, t}(Q) \: = \: \ms{rate}_{\rm ot}(Q, \procs_{\rm M, ot}) \\
\hline
\end{array}\]
which coincides with the reciprocal of the average sojourn time associated with $Q$.

The behavioral equivalence can be defined in different ways depending on the interpretation of action
execution. We observe that the operational semantic rule for action prefix in the upper part of
Table~\ref{sosot} encodes an eager interpretation, because it permits no delay between the time at which an
action becomes enabled and the time at which the same action starts its execution. In contrast to the
integrated time case, a different interpretation of action execution cannot be encoded in the operational
semantic rules because time can solve choices due to the adoption of the race policy.

Following the durationless setting defined in~\cite{Cor}, an additional operational semantic rule of the
form:
\cws{0}{a \, . \, Q \arrow{t}{} a \, . \, Q \hspace{0.5cm} \forall t \in \natns}
has to be introduced to manage lazy actions in a deterministically timed process calculus, with $t$ being an
arbitrary delay between the time at which the action becomes enabled and the time at which the action can
start its execution. In the maximal progress case, the additional rule is applied only when $a \in
\ms{Name}_{\rm v}$ because $\tau$-actions cannot let time pass. The effect of the additional rule is that a
process term can let time pass iff so can all the actions it enables. This is a consequence of some of the
operational semantic rules for binary operators, which are of the form:
\cws{0}{{\infr{Q_{1} \arrow{t}{} Q'_{1} \hspace{0.8cm} Q_{2} \arrow{t}{} Q'_{2}}{Q_{1} + Q_{2} \arrow{t}{}
Q'_{1} + Q'_{2}}} \hspace{1.5cm} {\infr{Q_{1} \arrow{t}{} Q'_{1} \hspace{0.8cm} Q_{2} \arrow{t}{}
Q'_{2}}{Q_{1} \pco{S} Q_{2} \arrow{t}{} Q'_{1} \pco{S} Q'_{2}}}}
and hence formalize the fact that time does not solve choices.

The analogous additional operational semantic rule for handling lazy actions in a Markovian framework would
be of the form:
\cws{0}{a \, . \, Q \arrow{\lambda}{} a \, . \, Q \hspace{0.5cm} \forall \lambda \in \realns_{> 0}}
However, the resulting exponentially timed selfloops would have no impact on the underlying continuous-time
Markov chain, as already discussed at the end of Sect.~\ref{discussit}. Most importantly, the additional
rule would not produce the desired effect, because in a Markovian framework time can solve choices due to
the adoption of the race policy and therefore rules like those above for alternative and parallel
composition in the deterministically timed case are not appropriate in the stochastically timed case.

The desired effect is instead obtained by encoding the three different interpretations of action execution
into three different variants of orthogonal-time Markovian bisimilarity. All of them work like classical
bisimilarity for action execution. As regards time passing, the exit rate comparison is performed: only for
pairs of terms that cannot execute any action under eagerness; for all pairs of terms under laziness; only
for pairs of terms that cannot execute any $\tau$-action under maximal progress.

	\begin{definition}

An equivalence relation $\calb$ over $\procs_{\rm M, ot}$ is an eager/lazy/maximal-progress orthogonal-time
Markovian bisimulation iff, whenever $(Q_{1}, Q_{2}) \in \calb$, then:

		\begin{itemize}

\item For all action names $a \in \ms{Name}$:

			\begin{itemize}

\item Whenever $Q_{1} \arrow{a}{} Q'_{1}$, then $Q_{2} \arrow{a}{} Q'_{2}$ with $(Q'_{1}, Q'_{2}) \in
\calb$.

\item Whenever $Q_{2} \arrow{a}{} Q'_{2}$, then $Q_{1} \arrow{a}{} Q'_{1}$ with $(Q'_{1}, Q'_{2}) \in
\calb$.

			\end{itemize}

\item For all equivalence classes $D \in \procs_{\rm M, ot} / \calb$:
\cws{0}{\hspace*{-0.7cm} \ms{rate}_{\rm ot}(Q_{1}, D) \: = \: \ms{rate}_{\rm ot}(Q_{2}, D)}
whenever:

			\begin{itemize}

\item $Q_{1}$ and $Q_{2}$ cannot perform any action (eagerness).

\item $Q_{1}$ and $Q_{2}$ are arbitrary (laziness).

\item $Q_{1}$ and $Q_{2}$ cannot perform any $\tau$-action (maximal progress).

			\end{itemize}

		\end{itemize}

\noindent
Eager/lazy/maximal-progress orthogonal-time Markovian bisimilarity $\sbis{\rm MB, ot, e}$/$\sbis{\rm MB, ot,
l}$/$\sbis{\rm MB, ot, mp}$ is the union of all the eager/lazy/maximal-progress orthogonal-time Markovian
bisimulations.
\fullbox

	\end{definition}

It turns out $\sbis{\rm MB, ot, l} \, \subset \, \sbis{\rm MB, ot, mp} \, \subset \, \sbis{\rm MB, ot, e}$.
In contrast to $\sbis{\rm MB, ot, e}$, which is not a congruence with respect to parallel composition,
$\sbis{\rm MB, ot, l}$ and $\sbis{\rm MB, ot, mp}$ can be shown to be congruences with respect to all the
operators of OTMPC as well as recursion and to have a sound and complete axiomatization over nonrecursive
process terms including typical laws like associativity, commutativity, and neutral element for the
alternative composition operator, the expansion law for the parallel composition operator, and distributive
laws for hiding and relabeling with respect to alternative composition. In particular, the characterizing
laws of $\sbis{\rm MB, ot, mp}$, which has been proposed and studied in~\cite{Her}, formalize the usual
idempotency of the alternative composition operator for action execution, the race policy for time passing,
and maximal progress:
\cws{10}{\begin{array}{rcl}
a \, . \, Q + a \, . \, Q & \!\!\! \sbis{\rm MB, ot, mp} \!\!\! & a \, . \, Q \\
(\lambda_{1}) \, . \, Q + (\lambda_{2}) \, . \, Q & \!\!\! \sbis{\rm MB, ot, mp} \!\!\! & (\lambda_{1} +
\lambda_{2}) \, . \, Q \\
\tau \, . \, Q + (\lambda) \, . \, Q' & \!\!\! \sbis{\rm MB, ot, mp} \!\!\! & \tau \, . \, Q \\
\end{array}}

%%%%%%%%%%%%%%%%%%%%%%%%%%%%%%%%%%%%%%%%%%%%%%%%%%%%%%%%%%%%%%%%%
%
%
\section{Encoding ITMPC into OTMPC}\label{encoding}
%
%
%%%%%%%%%%%%%%%%%%%%%%%%%%%%%%%%%%%%%%%%%%%%%%%%%%%%%%%%%%%%%%%%%

In this section, we show that a connection can be established between Markovian process calculi with
durational actions and Markovian process calculi with durationless actions. First, we single out the classes
of process terms of ITMPC and OTMPC for which a translation is possible under eagerness, laziness, and
maximal progress. Then, for each of the three interpretations of action execution, we formalize the encoding
of the related class of process terms of ITMPC into the related class of process terms of OTMPC and we prove
that it preserves the related bisimulation-based behavioral equivalence of the considered process terms.

%%%%%%%%%%%%%%%%%%%%%%%%%%%%%%%%%%%%%%%%%%%%%%%%%%%%%%%%%%%%%%%%%
%
\subsection{Classes of Process Terms}
%
%%%%%%%%%%%%%%%%%%%%%%%%%%%%%%%%%%%%%%%%%%%%%%%%%%%%%%%%%%%%%%%%%

In the deterministically timed case, the basic rule of the translating function defined in~\cite{Cor} maps
$\lap a, n \rap . P$ to $a \, . \, (n) \, . \, Q$ or $(n) \, . \, a \, . \, Q$ depending on whether the
starting time -- first option -- or the completion time \linebreak -- second option -- of the execution of
timed actions is observed, respectively, where $n$ is a fixed duration and $Q$ is the translation of $P$. As
a consequence, a process term like $\lap a_{1}, n_{1} \rap . P_{1} + \lap a_{2}, n_{2} \rap . P_{2}$ is
mapped to $a_{1} \, . \, (n_{1}) \, . \, Q_{1} + a_{2} \, . \, (n_{2}) \, . \, Q_{2}$ or $(n_{1}) \, . \,
a_{1} \, . \, Q_{1} + (n_{2}) \, . \, a_{2} \, . \, Q_{2}$, with the choice being nondeterministic in all
the three terms as time does not solve choices in this setting.

On the basis of the observation made in Sect.~\ref{intro}, the first option would not work in a
stochastically timed setting. In fact, the choice in a process term like $\lap a_{1}, \lambda_{1} \rap .
P_{1} + \lap a_{2}, \lambda_{2} \rap . P_{2}$ is probabilistic, whereas the choice in the corresponding
process term $a_{1} \, . \, (\lambda_{1}) \, . \, Q_{1} + a_{2} \, . \, (\lambda_{2}) \, . \, Q_{2}$ would
be nondeterministic. As a consequence, the basic rule of the translating function from ITMPC to OTMPC should
map $\lap a, \lambda \rap . P$ to $(\lambda) \, . \, a \, . \, Q$ -- with $Q$ being the translation of $P$
-- so that a process term like $\lap a_{1}, \lambda_{1} \rap . P_{1} + \lap a_{2}, \lambda_{2} \rap . P_{2}$
is mapped to $(\lambda_{1}) \, . \, a_{1} \, . \, Q_{1} + (\lambda_{2}) \, . \, a_{2} \, . \, Q_{2}$ -- with
the choice being probabilistic in both terms. In other words, ITMPC process terms can be translated only
into OTMPC process terms that do not contain nondeterministic choices. In the following, we denote by
$\procs_{\rm M, ot, nnd}$ the set of process terms of $\procs_{\rm M, ot}$ with no nondeterministic choices.

Selecting the appropriate order for action execution and time passing is not enough to achieve an encoding
that preserves the bisimulation-based behavioral equivalence of process terms. In fact, consider the
following ITMPC process terms:
\cws{0}{\begin{array}{rcl}
P_{1} & \!\!\! \equiv \!\!\! & \lap a, \lambda \rap . \nil \, \pco{\emptyset} \, \lap b, \mu \rap . \nil \\
P_{2} & \!\!\! \equiv \!\!\! & \lap a, \lambda \rap . \lap b, \mu \rap . \nil + \lap b, \mu \rap . \lap a,
\lambda \rap . \nil \\
\end{array}}
and the corresponding OTMPC process terms:
\cws{0}{\begin{array}{rcl}
Q_{1} & \!\!\! \equiv \!\!\! & (\lambda) \, . \, a \, . \, \nil \, \pco{\emptyset} \, (\mu) \, . \, b \, .
\, \nil \\
Q_{2} & \!\!\! \equiv \!\!\! & (\lambda) \, . \, a \, . \, (\mu) \, . \, b \, . \, \nil + (\mu) \, . \, b \,
. \, (\lambda) \, . \, a \, . \, \nil \\
\end{array}}
It turns out that $P_{1} \sbis{\rm MB, it} P_{2}$ because their underlying labeled multitransition systems
are isomorphic. By contrast, $Q_{1} \not\sbis{\rm MB, ot, l} Q_{2}$ because $\lsp Q_{1} \rsp_{\rm M, ot}$
contains states having both action transitions and time transitions due to interleaving, whereas this is not
the case with $\lsp Q_{2} \rsp_{\rm M, ot}$ as can be seen below: \\[0.1cm]
\centerline{\epsfbox{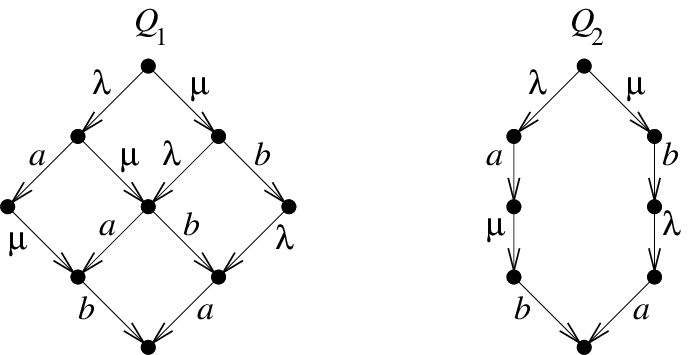}}
This shows that a translation of ITMPC into OTMPC is possible under laziness only for sequential process
terms, i.e., process terms that do not contain any occurrence of the parallel composition operator. In the
following, we denote by $\procs_{\rm M, it, seq}$ (resp.\ $\procs_{\rm M, ot, nnd, seq}$) the set of
sequential process terms of $\procs_{\rm M, it}$ (resp.\ $\procs_{\rm M, ot, nnd}$).

On the other hand, we have $Q_{1} \sbis{\rm MB, ot, e} Q_{2}$ because under eagerness action execution
always takes precedence over time passing, so that the central state of $\lsp Q_{1} \rsp_{\rm M, ot}$ and
its incoming transitions can be ignored when checking for orthogonal-time Markovian bisimilarity. Similarly,
we have $Q_{1} \sbis{\rm MB, ot, mp} Q_{2}$ whenever $a = \tau = b$. Should this not be the case, it would
be enough to add a $\tau$-selfloop to every state of $\lsp Q_{1} \rsp_{\rm M, ot}$ enabling an action. In
other words, under maximal progress the basic rule of the translating function from ITMPC to OTMPC should
map $\lap a, \lambda \rap . P$ to $(\lambda) \, . \, \textrm{rec} \, Z : (\tau \, . \, Z + a \, . \, Q)$,
where $Q$ is the translation of $P$ and $Z$ does not occur free in $Q$. Note that by doing so we reintroduce
nondeterministic choices in a controlled way. In the following, we denote by $\procs_{\rm M, ot, cnd}$ the
set of process terms of $\procs_{\rm M, ot}$ with controlled nondeterministic choices.

We conclude by showing another issue related to the preservation of the bisimulation-based behavioral
equivalence of process terms. Consider the following ITMPC process terms:
\cws{0}{\begin{array}{rcl}
P_{3} & \!\!\! \equiv \!\!\! & \lap a, \lambda \rap . \nil \\
P_{4} & \!\!\! \equiv \!\!\! & \lap a, \lambda \rap . \nil + \lap b, \mu \rap . \nil \, \pco{\{ b \}} \,
\nil \\
\end{array}}
and the corresponding OTMPC process terms:
\cws{0}{\begin{array}{rcl}
Q_{3} & \!\!\! \equiv \!\!\! & (\lambda) \, . \, a \, . \, \nil \\
Q_{4} & \!\!\! \equiv \!\!\! & (\lambda) \, . \, a \, . \, \nil + (\mu) \, . \, b \, . \, \nil \, \pco{\{ b
\}} \, \nil \\
\end{array}}
It turns out that $P_{3} \sbis{\rm MB, it} P_{4}$ because their underlying labeled multitransition systems
are isomorphic. By contrast, $Q_{3} \not\sbis{\rm MB, ot, e} Q_{4}$ and $Q_{3} \not\sbis{\rm MB, ot, mp}
Q_{4}$ as can be seen from their underlying labeled multitransition systems shown below: \\[0.1cm]
\centerline{\epsfbox{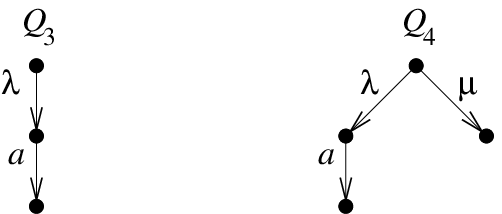}}
Here, the problem is that $\lsp Q_{4} \rsp_{\rm M, ot}$ has a spurious deadlock state deriving from the need
of encoding every exponentially timed action as its rate followed by its name. This problem can only arise
in the presence of restrictions on the actions that can be executed. According to the syntax of ITMPC, this
can only happen in the presence of occurrences of the parallel composition operator whose synchronization
set is not empty. Therefore, a translation of ITMPC into OTMPC is possible under eagerness and maximal
progress only for synchronization-free process terms. In the following, we denote by $\procs_{\rm M, it,
sf}$ (resp.\ $\procs_{\rm M, ot, nnd, sf}$/$\procs_{\rm M, ot, cnd, sf}$) the set of synchronization-free
process terms of $\procs_{\rm M, it}$ (resp.\ $\procs_{\rm M, ot, nnd}$/$\procs_{\rm M, ot, cnd}$).

%%%%%%%%%%%%%%%%%%%%%%%%%%%%%%%%%%%%%%%%%%%%%%%%%%%%%%%%%%%%%%%%%
%
\subsection{Translating Function for Laziness}\label{transl}
%
%%%%%%%%%%%%%%%%%%%%%%%%%%%%%%%%%%%%%%%%%%%%%%%%%%%%%%%%%%%%%%%%%

The function $\Gamma_{\rm l} : \procs_{\rm M, it, seq} \rightarrow \procs_{\rm M, ot, nnd, seq}$ encoding
ITMPC into OTMPC under laziness is defined by structural induction as follows:
\[\begin{array}{|rcl|}
\hline
\Gamma_{\rm l} \lsp \nil \rsp & \!\!\! = \!\!\! & \nil \\
\Gamma_{\rm l} \lsp \lap a, \lambda \rap . P \rsp & \!\!\! = \!\!\! & (\lambda) \, . \, a \, . \,
\Gamma_{\rm l} \lsp P \rsp \\
\Gamma_{\rm l} \lsp P_{1} + P_{2} \rsp & \!\!\! = \!\!\! & \Gamma_{\rm l} \lsp P_{1} \rsp + \Gamma_{\rm l}
\lsp P_{2} \rsp \\
\Gamma_{\rm l} \lsp P / H \rsp & \!\!\! = \!\!\! & \Gamma_{\rm l} \lsp P \rsp / H \\
\Gamma_{\rm l} \lsp P[\varphi] \rsp & \!\!\! = \!\!\! & \Gamma_{\rm l} \lsp P \rsp [\varphi] \\
\Gamma_{\rm l} \lsp X \rsp & \!\!\! = \!\!\! & X \\
\Gamma_{\rm l} \lsp \textrm{rec} \, X : P \rsp & \!\!\! = \!\!\! & \textrm{rec} \, X : \Gamma_{\rm l} \lsp P
\rsp \\
\hline
\end{array}\]

We now prove that $\Gamma_{\rm l}$ preserves the bisimulation-based behavioral equivalence of the considered
process terms by first demonstrating some useful properties of $\Gamma_{\rm l}$, among which the fact that
every ITMPC sequential process term and its $\Gamma_{\rm l}$-translation into OTMPC possess the same total
exit rate.

	\begin{lemma}\label{reclemmal}

Let $P \in \calpl_{\rm M, it, seq}$, $\textrm{rec} \, X : \hat{P} \in \procs_{\rm M, it, seq}$, and $Y \in
\ms{Var}$. Then:
\cws{10}{\Gamma_{\rm l} \lsp P \{ \textrm{rec} \, X : \hat{P} \hookrightarrow Y \} \rsp \: = \: \Gamma_{\rm
l} \lsp P \rsp \{ \textrm{rec} \, X : \Gamma_{\rm l} \lsp \hat{P} \rsp \hookrightarrow Y \}}
\fullbox

	\end{lemma}

	\begin{lemma}\label{terlemmal}

Let $P \in \procs_{\rm M, it, seq}$. Then $\Gamma_{\rm l} \lsp P \rsp$ cannot perform any action and:
\cws{10}{\ms{rate}_{\rm it, t}(P) \: = \: \ms{rate}_{\rm ot, t}(\Gamma_{\rm l} \lsp P \rsp)}
\fullbox

	\end{lemma}

	\begin{lemma}\label{translemmal}

Let $P \in \procs_{\rm M, it, seq}$. Then $P \arrow{a, \lambda}{\rm M, it} P'$ iff $\Gamma_{\rm l} \lsp P
\rsp \arrow{\lambda}{\rm M} Q$ with the only transition of \linebreak $Q \in \procs_{\rm M, ot, nnd, seq}$
being $Q \arrow{a}{} \Gamma_{\rm l} \lsp P' \rsp$.
\fullbox

	\end{lemma}

	\begin{theorem}\label{transthml}

Let $P_{1}, P_{2} \in \procs_{\rm M, it, seq}$. Then:
\cws{10}{P_{1} \sbis{\rm MB, it} P_{2} \: \Longleftrightarrow \: \Gamma_{\rm l} \lsp P_{1} \rsp \sbis{\rm
MB, ot, l} \Gamma_{\rm l} \lsp P_{2} \rsp}
\fullbox

	\end{theorem}

%%%%%%%%%%%%%%%%%%%%%%%%%%%%%%%%%%%%%%%%%%%%%%%%%%%%%%%%%%%%%%%%%
%
\subsection{Translating Function for Eagerness}\label{transe}
%
%%%%%%%%%%%%%%%%%%%%%%%%%%%%%%%%%%%%%%%%%%%%%%%%%%%%%%%%%%%%%%%%%

The function $\Gamma_{\rm e} : \procs_{\rm M, it, sf} \rightarrow \procs_{\rm M, ot, nnd, sf}$ encoding
ITMPC into OTMPC under eagerness is defined by structural induction as follows:
\[\begin{array}{|rcl|}
\hline
\Gamma_{\rm e} \lsp \nil \rsp & \!\!\! = \!\!\! & \nil \\
\Gamma_{\rm e} \lsp \lap a, \lambda \rap . P \rsp & \!\!\! = \!\!\! & (\lambda) \, . \, a \, . \,
\Gamma_{\rm e} \lsp P \rsp \\
\Gamma_{\rm e} \lsp P_{1} + P_{2} \rsp & \!\!\! = \!\!\! & \Gamma_{\rm e} \lsp P_{1} \rsp + \Gamma_{\rm e}
\lsp P_{2} \rsp \\
\Gamma_{\rm e} \lsp P_{1} \pco{\emptyset} P_{2} \rsp & \!\!\! = \!\!\! & \Gamma_{\rm e} \lsp P_{1} \rsp
\pco{\emptyset} \Gamma_{\rm e} \lsp P_{2} \rsp \\
\Gamma_{\rm e} \lsp P / H \rsp & \!\!\! = \!\!\! & \Gamma_{\rm e} \lsp P \rsp / H \\
\Gamma_{\rm e} \lsp P[\varphi] \rsp & \!\!\! = \!\!\! & \Gamma_{\rm e} \lsp P \rsp [\varphi] \\
\Gamma_{\rm e} \lsp X \rsp & \!\!\! = \!\!\! & X \\
\Gamma_{\rm e} \lsp \textrm{rec} \, X : P \rsp & \!\!\! = \!\!\! & \textrm{rec} \, X : \Gamma_{\rm e} \lsp P
\rsp \\
\hline
\end{array}\]
where the only difference with respect to $\Gamma_{\rm l}$ is the presence of a clause for parallel
composition.

%ZZZ
%Similar to $\Gamma_{\rm l}$, we now prove that $\Gamma_{\rm e}$ preserves the bisimulation-based behavioral
%equivalence of the considered process terms by first demonstrating some useful properties of $\Gamma_{\rm
%e}$.

	\begin{lemma}\label{reclemmae}

Let $P \in \calpl_{\rm M, it, sf}$, $\textrm{rec} \, X : \hat{P} \in \procs_{\rm M, it, sf}$, and $Y \in
\ms{Var}$. Then:
\cws{10}{\Gamma_{\rm e} \lsp P \{ \textrm{rec} \, X : \hat{P} \hookrightarrow Y \} \rsp \: = \: \Gamma_{\rm
e} \lsp P \rsp \{ \textrm{rec} \, X : \Gamma_{\rm e} \lsp \hat{P} \rsp \hookrightarrow Y \}}
\fullbox

	\end{lemma}

	\begin{lemma}\label{terlemmae}

Let $P \in \procs_{\rm M, it, sf}$. Then $\Gamma_{\rm e} \lsp P \rsp$ cannot perform any action and:
\cws{10}{\ms{rate}_{\rm it, t}(P) \: = \: \ms{rate}_{\rm ot, t}(\Gamma_{\rm e} \lsp P \rsp)}
\fullbox

	\end{lemma}

	\begin{lemma}\label{translemmae}

Let $P \in \procs_{\rm M, it, sf}$. Then $P \arrow{a, \lambda}{\rm M, it} P'$ iff $\Gamma_{\rm e} \lsp P
\rsp \arrow{\lambda}{\rm M} Q$ with the only action transition of $Q \in \procs_{\rm M, ot, nnd, sf}$ being
$Q \arrow{a}{} \Gamma_{\rm e} \lsp P' \rsp$.
\fullbox

	\end{lemma}

	\begin{theorem}\label{transthme}

Let $P_{1}, P_{2} \in \procs_{\rm M, it, sf}$. Then:
\cws{10}{P_{1} \sbis{\rm MB, it} P_{2} \: \Longleftrightarrow \: \Gamma_{\rm e} \lsp P_{1} \rsp \sbis{\rm
MB, ot, e} \Gamma_{\rm e} \lsp P_{2} \rsp}
\fullbox

	\end{theorem}

%%%%%%%%%%%%%%%%%%%%%%%%%%%%%%%%%%%%%%%%%%%%%%%%%%%%%%%%%%%%%%%%%
%
\subsection{Translating Function for Maximal Progress}\label{transmp}
%
%%%%%%%%%%%%%%%%%%%%%%%%%%%%%%%%%%%%%%%%%%%%%%%%%%%%%%%%%%%%%%%%%

The function $\Gamma_{\rm mp} : \procs_{\rm M, it, sf} \rightarrow \procs_{\rm M, ot, cnd, sf}$ encoding
ITMPC into OTMPC under maximal progress is defined by structural induction as follows:
\[\begin{array}{|rcll|}
\hline
\Gamma_{\rm mp} \lsp \nil \rsp & \!\!\! = \!\!\! & \nil & \\
\Gamma_{\rm mp} \lsp \lap a, \lambda \rap . P \rsp & \!\!\! = \!\!\! & (\lambda) \, . \, \textrm{rec} \, Z :
(\tau \, . \, Z + a \, . \, \Gamma_{\rm mp} \lsp P \rsp) & \hspace{0.5cm} \textrm{$Z$ not free in $P$} \\
\Gamma_{\rm mp} \lsp P_{1} + P_{2} \rsp & \!\!\! = \!\!\! & \Gamma_{\rm mp} \lsp P_{1} \rsp + \Gamma_{\rm
mp} \lsp P_{2} \rsp & \\
\Gamma_{\rm mp} \lsp P_{1} \pco{\emptyset} P_{2} \rsp & \!\!\! = \!\!\! & \Gamma_{\rm mp} \lsp P_{1} \rsp
\pco{\emptyset} \Gamma_{\rm mp} \lsp P_{2} \rsp & \\
\Gamma_{\rm mp} \lsp P / H \rsp & \!\!\! = \!\!\! & \Gamma_{\rm mp} \lsp P \rsp / H & \\
\Gamma_{\rm mp} \lsp P[\varphi] \rsp & \!\!\! = \!\!\! & \Gamma_{\rm mp} \lsp P \rsp [\varphi] & \\
\Gamma_{\rm mp} \lsp X \rsp & \!\!\! = \!\!\! & X & \\
\Gamma_{\rm mp} \lsp \textrm{rec} \, X : P \rsp & \!\!\! = \!\!\! & \textrm{rec} \, X : \Gamma_{\rm mp} \lsp
P \rsp & \\
\hline
\end{array}\]
where the only difference with respect to $\Gamma_{\rm e}$ is the clause for action prefix, which introduces
$\tau$-selfloops.

%ZZZ
%Similar to $\Gamma_{\rm e}$, we now prove that $\Gamma_{\rm mp}$ preserves the bisimulation-based behavioral
%equivalence of the considered process terms by first demonstrating some useful properties of $\Gamma_{\rm
%mp}$.

	\begin{lemma}\label{reclemmamp}

Let $P \in \calpl_{\rm M, it, sf}$, $\textrm{rec} \, X : \hat{P} \in \procs_{\rm M, it, sf}$, and $Y \in
\ms{Var}$. Then:
\cws{10}{\Gamma_{\rm mp} \lsp P \{ \textrm{rec} \, X : \hat{P} \hookrightarrow Y \} \rsp \: = \: \Gamma_{\rm
mp} \lsp P \rsp \{ \textrm{rec} \, X : \Gamma_{\rm mp} \lsp \hat{P} \rsp \hookrightarrow Y \}}
\fullbox

	\end{lemma}

	\begin{lemma}\label{terlemmamp}

Let $P \in \procs_{\rm M, it, sf}$. Then $\Gamma_{\rm mp} \lsp P \rsp$ cannot perform any action and:
\cws{10}{\ms{rate}_{\rm it, t}(P) \: = \: \ms{rate}_{\rm ot, t}(\Gamma_{\rm mp} \lsp P \rsp)}
\fullbox

	\end{lemma}

	\begin{lemma}\label{translemmamp}

Let $P \in \procs_{\rm M, it, sf}$. Then $P \arrow{a, \lambda}{\rm M, it} P'$ iff $\Gamma_{\rm mp} \lsp P
\rsp \arrow{\lambda}{\rm M} Q$ with the only action transitions of $Q \in \procs_{\rm M, ot, cnd, sf}$ being
$Q \arrow{\tau}{} Q$ and $Q \arrow{a}{} \Gamma_{\rm mp} \lsp P' \rsp$.
\fullbox

	\end{lemma}

	\begin{theorem}\label{transthmmp}

Let $P_{1}, P_{2} \in \procs_{\rm M, it, sf}$. Then:
\cws{10}{P_{1} \sbis{\rm MB, it} P_{2} \: \Longleftrightarrow \: \Gamma_{\rm mp} \lsp P_{1} \rsp \sbis{\rm
MB, ot, mp} \Gamma_{\rm mp} \lsp P_{2} \rsp}
\fullbox

	\end{theorem}

%%%%%%%%%%%%%%%%%%%%%%%%%%%%%%%%%%%%%%%%%%%%%%%%%%%%%%%%%%%%%%%%%
%
%
\section{Conclusion}\label{concl}
%
%
%%%%%%%%%%%%%%%%%%%%%%%%%%%%%%%%%%%%%%%%%%%%%%%%%%%%%%%%%%%%%%%%%

In this paper, we have shown that durational actions and durationless actions are not irreconcilable even in
a stochastically timed setting, because we have exhibited suitable semantics-preserving mappings from an
integrated-time Markovian process calculus to an orthogonal-time Markovian under eagerness, laziness, and
maximal progress. The restrictions on the three mappings emphasize synchronization disciplines and choice
resolutions as the only features distinguishing between the two considered calculi.

We have also highlighted a number of differences with respect to the deterministically timed setting
examined in~\cite{Cor}. Firstly, due to the adoption of the race policy, time solves choices and hence any
exponentially timed action must be translated into an exponentially distributed delay followed by an
instantaneous action, rather than the opposite. Secondly, in the integrated-time case the memoryless
property of exponential distributions blurs the distinction among eagerness, laziness, and maximal progress.
Thirdly, since time solve choices, in the orthogonal-time case the three interpretations of action execution
must be formalized through as many variants of the behavioral equivalence, rather than in the operational
semantic rules. Fourthly, the mapping for laziness is limited to sequential process terms, rather than being
applicable in general. Sixtly, the mapping for maximal progress is limited to synchronization-free process
terms and needs the introduction of $\tau$-selfloops, rather than being applicable in general. Seventhly,
the three mappings constrain the amount of nondeterminism in the resulting process terms, rather than
admitting full nondeterminism.

Orthogonal-time Markovian process calculi turn out to be more expressive as they can represent both
probabilistic and nondeterministic choices as well as more natural forms of synchronization. Nevertheless,
integrated-time Markovian process calculi should not be neglected. Firstly, they are in general more
appropriate for modeling purposes, because it is more natural to think of an action as having a duration
rather than expressing a delay followed by an action name. Secondly, unlike orthogonal-time Markovian
process calculi they do not incur in spurious deadlock states. Thirdly, they tend to produce system
descriptions with no more than half of the states that would result from descriptions of the same systems
expressed in orthogonal-time Markovian process calculi.

\bigskip
\noindent
\textbf{Acknowledgment}: This work has been funded by MIUR-PRIN project \textit{PaCo -- Performability-Aware
Computing: Logics, Models, and Languages}.

%%%%%%%%%%%%%%%%%%%%%%%%%%%%%%%%%%%%%%%%%%%%%%%%%%%%%%%%%%%%%%%%%
%                                                               %
%                                                               %
% References                                                    %
%                                                               %
%                                                               %
%%%%%%%%%%%%%%%%%%%%%%%%%%%%%%%%%%%%%%%%%%%%%%%%%%%%%%%%%%%%%%%%%
\bibliographystyle{eptcs}

\end{document}

%% file: Commands.tex
\def\ms#1{\null\ifmmode\mathord{\mathcode`-="702D\it #1\mathcode`\-="2200}%
	\else$\mathord{\mathcode`-="702D\it #1\mathcode`\-="2200}$\fi}

\newcommand{\cws}[2]
	{\\ \centerline{$#2$} \\[-#1pt]}

\newlength{\spacelen}

\newcommand{\lap}
	{\mbox{$<$}}
\newcommand{\rap}
	{\mbox{$>$}}

\newcommand{\lsp}
	{[ \! [}
\newcommand{\rsp}
	{] \! ]}

\newcommand{\lmp}
	{\{ \! | \,}
\newcommand{\rmp}
	{\, | \! \}}

\newcommand{\calb}
        {\mathcal{B}}

\newcommand{\calpl}
        {\mathcal{PL}}

\newcommand{\natns}
	{\mathbb{N}}

\newcommand{\realns}
	{\mathbb{R}}

\newcommand{\procs}
	{\mathbb{P}}

\newcommand{\arrow}[2]
        {\, {\auxarrow\limits^{#1}}_{#2} \,}
\newcommand{\auxarrow}
	{\mathop{- \!\! - \!\!\!\! \longrightarrow}}

\newcommand{\nil}
	{\underline 0}

\newcommand{\sbis}[1]
	{\sim_{#1}}

\newcommand{\pco}[1]
	{\mathop{\Vert_{#1}}}

\newcommand{\infr}[2]
	{\renewcommand{\arraystretch}{1.5}
	\begin{array}{c}
	#1\\
	\hline
	#2
	\end{array}}

\newcommand{\fullbox}
	{{\mbox{}\nolinebreak\hfill{$\rule{2mm}{2mm}$}}}

\newcommand{\empagr}
	{\textrm{EMPA$_{\rm gr}$}}

%% file: Environments.tex
\newtheorem{new_theorem}
	{Theorem}[section]

\newtheorem{new_definition}
	[new_theorem]{Definition}

\newtheorem{new_remark}
	[new_theorem]{Remark}

\newtheorem{new_example}
	[new_theorem]{Example}

\newtheorem{new_lemma}
	[new_theorem]{Lemma}

\newtheorem{new_proposition}
	[new_theorem]{Proposition}

\newtheorem{new_corollary}
	[new_theorem]{Corollary}

\newenvironment{definition}
	{\begin{new_definition}\rm}
	{\end{new_definition}}

\newenvironment{lemma}
	{\begin{new_lemma}\rm}
	{\end{new_lemma}}

\newenvironment{theorem}
	{\begin{new_theorem}\rm}
	{\end{new_theorem}}